\documentclass[aps,prl,preprint,floatfix,groupedaddress,showpacs]{revtex4}
\usepackage[dvips]{graphicx}
\usepackage{subfigure}

\begin{document}
\title{Frequency matching in light storage spectroscopy of atomic Raman transitions}

\author{Leon Karpa}
\affiliation{Institut f\"ur Angewandte Physik der Universit\"at Bonn, Wegelerstr. 8, 53115 Bonn, Germany}
\author{Gor Nikoghosyan}
\affiliation{Fachbereich Physik und Forschungszentrum OPTIMAS,
Technische Universit\"at Kaiserslautern, 67663 Kaiserslautern,
Germany }
\author{Frank Vewinger}
\affiliation{Institut f\"ur Angewandte Physik der
Universit\"at Bonn, Wegelerstr. 8, 53115 Bonn, Germany}

\author{Michael Fleischhauer}
\affiliation{Fachbereich Physik, Technische Universit\"at Kaiserslautern, 67663 Kaiserslautern, Germany }
\author{Martin Weitz}
\affiliation{Institut f\"ur Angewandte Physik der Universit\"at Bonn, Wegelerstr. 8, 53115 Bonn, Germany}

\date{\today}


\begin{abstract}
We investigate the storage of light in an atomic sample with a
lambda-type coupling scheme driven by optical fields at variable
two-photon detuning. In the presence of electromagnetically
induced transparency (EIT), light is stored and retrieved from the
sample by dynamically varying the group velocity. It is found that
for any two-photon detuning of the input light pulse within the
EIT transparency window, the carrier frequency of the retrieved
light pulse matches the two-photon resonance frequency with the
atomic ground state transition and the control field. This effect
which is not based on spectral filtering is investigated both
theoretically and experimentally. It can be used for high-speed
precision measurements of the two-photon resonance as employed
e.g. in optical magnetometry.
\end{abstract}


\maketitle

Optically dense atomic media can be made transparent by means of
destructive quantum interference of optical absorption amplitudes
\cite{10.1103/RevModPhys.77.633}. Media prepared under such
conditions called electromagnetically induced transparency (EIT)
are known to show interesting properties, such as extremely slow
group velocities
\cite{Nature.397.594-598(1999),PhysRevLett.82.5229,PhysRevLett.83.1767}.
Storage of light can be achieved by adiabatically reducing the
group velocity to zero allowing for reversible storage of photonic
information in an atomic ground state coherence. In recent
experiments also non-classical light states such as squeezed and
single-photon states have been reversibly stored in atomic media
\cite{appel:093602,Nature.409.490-493(2001),PhysRevLett.86.783,10.1038/nature04327}.
Light storage promises to be of importance in the field of quantum
information, where applications as the implementation of quantum
storage devices, gates and generation of photonic qubits have been
suggested \cite{Duan2001,Chaneliere2005,Vewinger2007}.
Electromagnetically induced transparency has been shown to enable
metrological applications like optical magnetometry. Due to the
narrow linewidths obtainable with dark resonances
\cite{Brandt1997}, precise magnetometers can be
realized \cite{PhysRevLett.69.1360,Katsoprinakis2006}. Recent
advances in this field have lead to the development of state of
the art magnetometers surpassing the precision of superconducting
quantum interference devices \cite{Kominis2003}. Recently it has
been demonstrated by some of the authors that the storage of light
in an atomic tripod system after retrieval allows for the
observation of an optical beat signal resonant to an atomic ground
state coherence \cite{Karpa2008a}.

Here we demonstrate that the observed resonance beating is a general feature of light storage
in optically dense ensembles of atoms with dark states.
Storage of light was performed by dynamically reducing the optical group velocity
in the dark state system to zero. After reaccelerating the stored light, a beating between
the released signal beam pulse and a control beam field is observed
which is shown to match the energetic splitting between the two ground state sublevels
for any two-photon detuning of the input pulse within the EIT transparency window.
The frequency matching effect is explained in terms of the polariton picture of EIT
\cite{PhysRevLett.84.5094,Mewes-PRA-2002}.

The observed synchronization effect allows for a novel method for
atomic spectroscopy between ground state sublevels, as is of
interest both in the field of atomic clocks and the measurements
of magnetic fields. Our experimental measurements indicate that
with light storage spectroscopy two-photon resonance frequencies
can be measured within a fraction of the time required for the
acquisition of a complete dark resonance spectrum. Furthermore,
the scheme is relatively robust against fluctuations of the
precise optical frequencies of the two incident light fields. As
long as difference frequency fluctuations do not exceed the
spectral width of the dark resonance, the retrieved frequency
difference remains locked to the two-photon resonance.

\begin{figure}[ht!]
  \begin{center}
    \includegraphics [height=4cm]{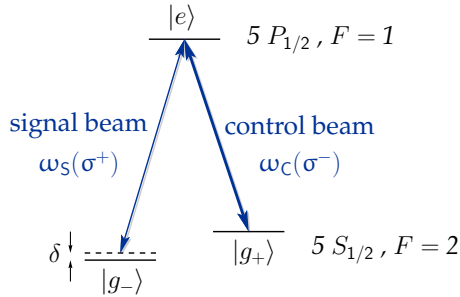}
  \caption{\label{fig:Lambda} Three-level scheme used for storage and retrieval of light. The levels $\left|g_{+}\right\rangle$
  and $\left|e\right\rangle$ and $\left|g_{-}\right\rangle$
  and $\left|e\right\rangle$ are coupled by a $\sigma^-$-polarized control field and a $\sigma^+$-polarized signal field with optical frequencies $\omega_c$ and $\omega_s$ respectively. The two-photon detuning is given by $\delta=\omega_s-\omega_c-2g_F\mu_B B$, where $\mu_B$ is the Bohr magnetron, $g_F$ the hyperfine g-factor and $B$ the magnetic field.}
  \end{center}
\end{figure}

Consider a medium consisting of atoms exhibiting a level structure
with two stable ground states $\left|g_-\right\rangle$,
$\left|g_+\right\rangle$ and one spontaneously decaying
electronically excited state $\left|e\right\rangle$, as depicted
in Fig.~\ref{fig:Lambda}.
The transition between the states $\left|g_+\right\rangle$ and
$\left|e\right\rangle$ is driven by a $\sigma^-$-polarized
"control" beam, which is assumed to be stronger than the
$\sigma^+$-polarized "signal" field coupling the states
$\left|g_-\right\rangle$ and $\left|e\right\rangle$. Under the
conditions of electromagnetically induced transparency the
propagation of light can be described in terms of a moving
quasiparticle, the so-called dark-state polariton, which is a
mixture of a photonic contribution and an atomic spin-wave
component \cite{PhysRevLett.84.5094}. For an ensemble of
three-level atoms, as shown in Fig. \ref{fig:Lambda}, interacting
with a strong control field and a weaker signal field, the group
velocity is proportional to the intensity of the control field,
and thus can be manipulated experimentally. The atom-light
quasiparticle even can be stopped by dynamically reducing the
optical field intensities to zero, which is referred to as
"storage of light"
\cite{Nature.409.490-493(2001),PhysRevLett.86.783}. During this
process information about the polarization state, phase and
amplitude of the incident light field is reversibly mapped onto a
spin wave coherence. After reactivation of the control beam, these
properties are again imprinted on the reaccelerated signal field.
It is remarkable that pure quantum states of light can in this way
be reversibly stored in the multiparticle system of an atomic gas
\cite{Lukin2000}.

\begin{figure}[ht!]
  \begin{center}
    \includegraphics [width=\columnwidth]{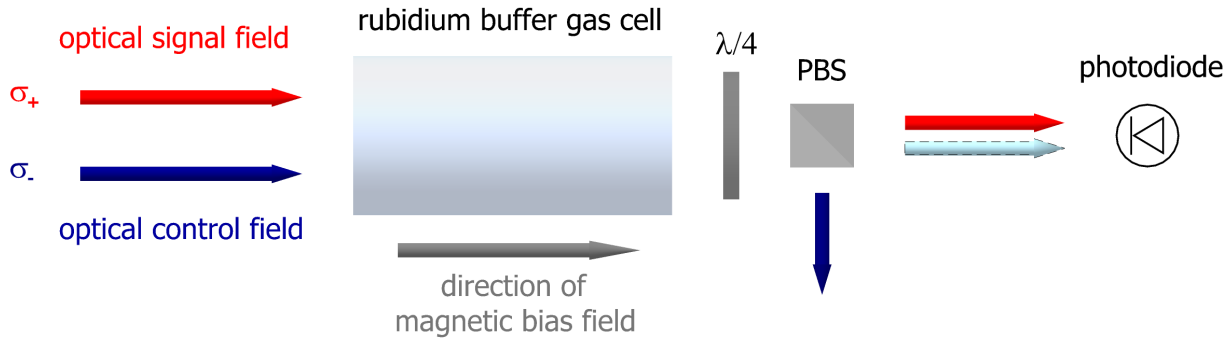}
  \caption{\label{fig:setup} Scheme of experimental setup.}
  \end{center}
\end{figure}

The apparatus used in our experiment (see Fig.~\ref{fig:setup}) is
a modified version of a previously described setup
\cite{2006NatPh.2.332K,Karpa2008a}. A grating stabilized diode
laser locked to the $F=2\longrightarrow F'=1$ component of the
rubidium $D1$-line near 795 nm is used a as laser source of both
the optical signal as well as the control field. The two beams
pass independent acousto-optical modulators to allow for a
variation of the difference frequency and intensity of the
individual beams. Inside a magnetically shielded region, a 50 mm
long rubidium vapor cell containing 10 torr neon buffer gas is
heated to approximately $80\,^{\circ}\mathrm{C}$. A magnetic bias
field directed along the optical beam axis is applied in order to
split up the ground state Zeeman components, which serve as states
$\left|g_{-}\right\rangle$ and $\left|g_{+}\right\rangle$ in the
experiment. The energetic separation of two neighboring ground
states is given by $g_F\mu_B B$, where $\mu_B$ is the Bohr
magneton, and the hyperfine g-factor equals $g_F=1/2$ for the used
transition.

After passing the acousto-optic modulators, the beam paths are
spatially overlapped and sent through a polarization maintaining
optical fiber. The beams are then expanded to 2 mm beam diameter
and sent through the rubidium cell with opposite circular
polarizations. In our experiments the control beam intensity was
always much stronger than the signal beam, so that most of the
relevant atoms were in the $F=2, m_F=-2$ ground state sublevel. In
this case, the states $\left|g_-\right\rangle$ and
$\left|g_+\right\rangle$ of the simplified three-level scheme
shown in Fig. \ref{fig:Lambda} correspond to the $m_F=-2$ and
$m_F=0$ Zeeman sub-levels of the $F=2$ ground state component
respectively. Under EIT conditions a dark coherent superposition
of these ground states is established by optical pumping. After
traversing the rubidium cell, with a $\lambda /4$-wave plate the
circular polarizations of signal and control beams can be
converted to opposite linear polarizations respectively and a
subsequent polarizing beamsplitter removes the control beam, so
that only the signal beam intensity is detected with an optical
photodiode. In order to observe an optical beating signal between
control and signal fields, a small portion of the control field is
required to reach the detector, which can be realized by a slight
rotation of the $\lambda/4$-plate.

\begin{figure}[h]
     \centering
        \subfigure[]{
           \label{fig:StoredLight}
           \includegraphics[height=4cm]{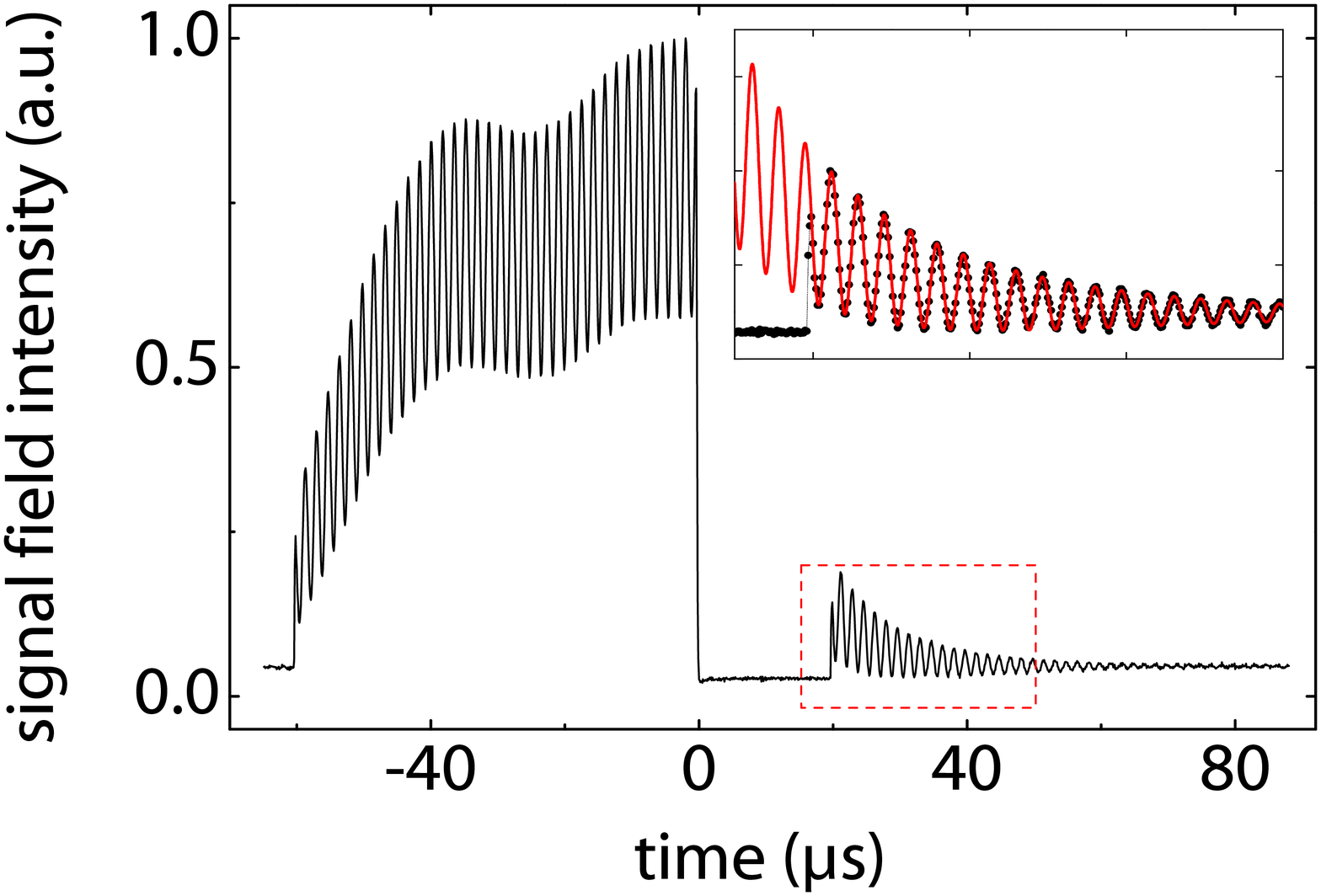}}
\hspace{0.5cm}
        \centering
     \subfigure[]{
           \label{fig:FvsB}
           \includegraphics[height=4cm]{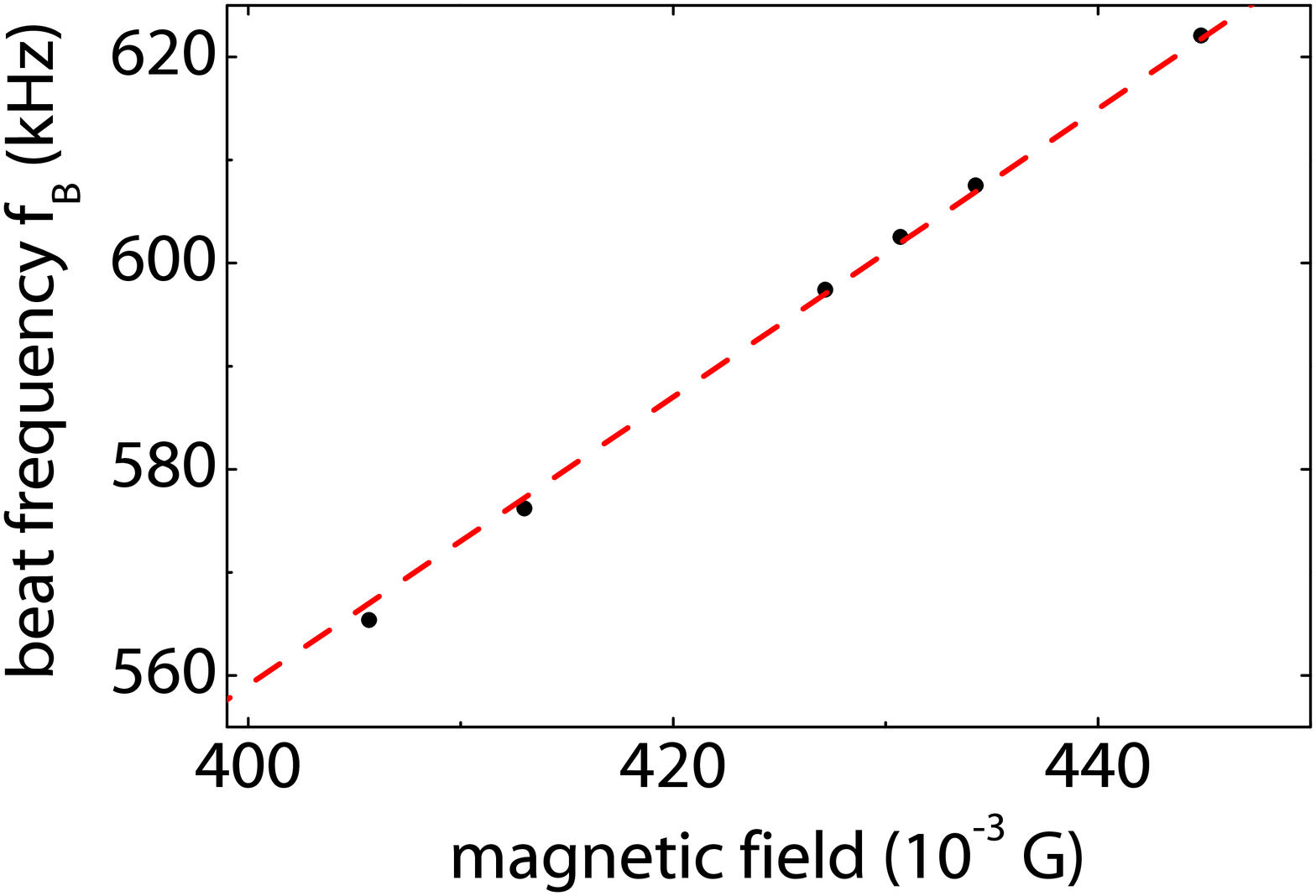}}
     \caption{\textbf{(a)} Typical stored light signal for a storage period of 20 $\mu s$. At time t=0 the control field and the signal field were switched off adiabatically, and at time $t\approx 20\,\mu$s only the control field was switched on, restoring the signal field. The inset shows a portion of the retrieved pulse fitted with a sinusoidally modulated function \cite{envelope}.
      \textbf{(b)} Measured beat frequency of the released signal light as a function of the applied magnetic bias field. All data points were recorded with the same value of the two-photon detuning of the initial signal field. The size of the error bars of the data points is smaller than the drawing size of the dots.}
     \label{fig:results}
\end{figure}

In initial experiments, we investigated both dark resonances and
the storage of light in our setup. Typical values for the optical
field power were 300 $\mu W$ for the control and 100 $\mu W$ for
the signal beam.
Subsequently the presence of a beating signal was
verified by irradiating the sample with initially rectangularly
shaped optical signal field pulses with a duration of
approximately 60 $\mu s$, with its falling edge set to the same
time as the falling edge of the control field. After a storage
period of typically 20 $\mu s$ only the control field was turned
on again triggering the retrieval of the signal field. The
detected signal, as shown in Fig.~\ref{fig:StoredLight}, showed
rapid oscillations on top of an exponentially decaying amplitude
signal due to the beat between the signal and control fields. The
oscillation frequency between the control field and the initial or
restored signal field was determined by fitting the recorded
signal with a sinusoidally modulated function.

\begin{figure}[h]
  \begin{center}
    \includegraphics [height=4cm]{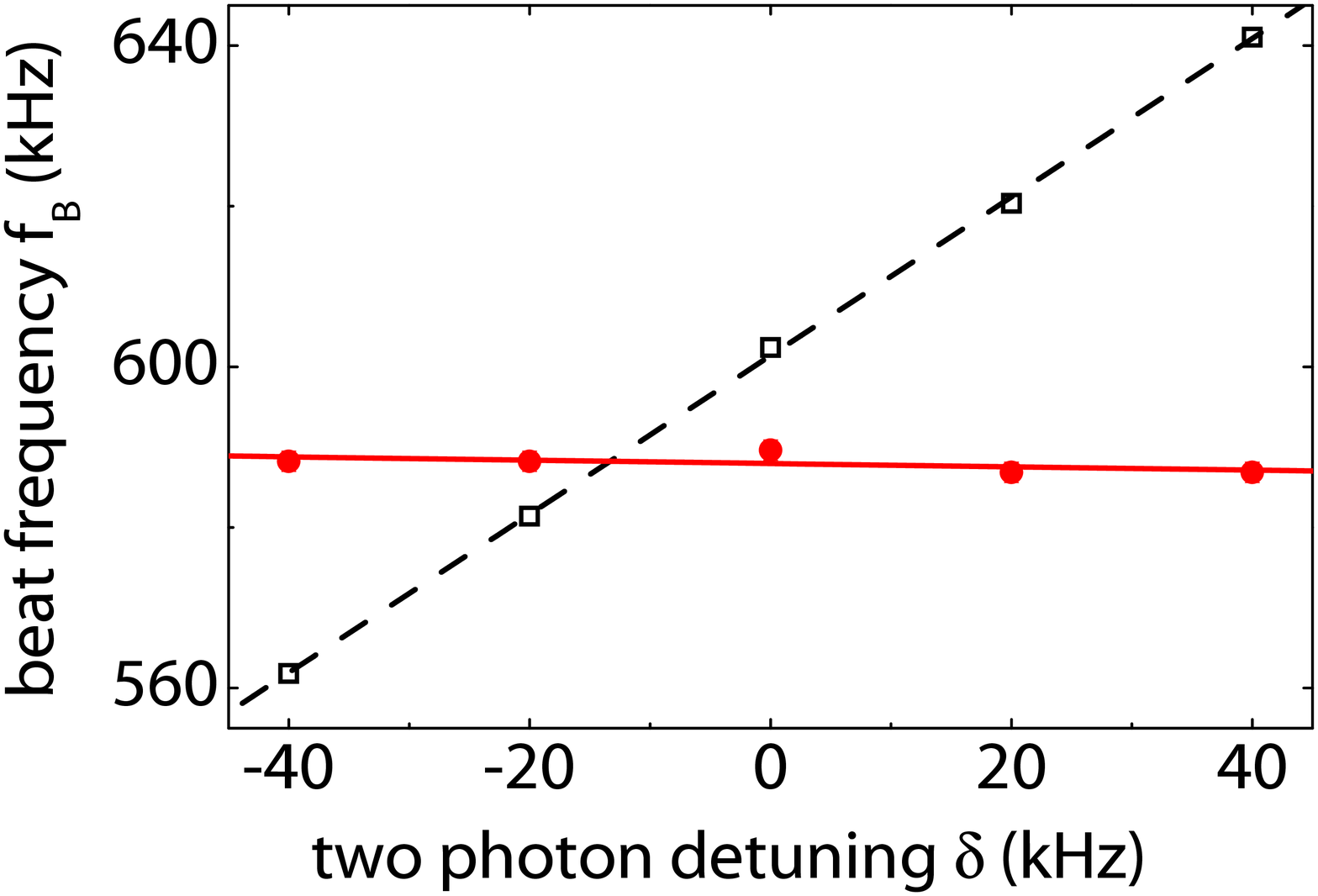}
  \caption{\label{fig:FvsDelta} Beat frequencies for different values of the two-photon detuning for the pulses before storage (open squares) and after the retrieval procedure (circles). For all shown data points the atomic sample was exposed to the same magnetic bias field. The slopes of the fitted linear curves are (within uncertainties calculated by the fitting routine) one for the input pulses (dashed line) and zero for the pulses detected after storage (solid line). The size of the error bars is within the drawing size of the data points.}
  \end{center}
\end{figure}

We have recorded such oscillation data for different values of the
magnetic bias field, while the values of the optical frequencies
of signal and control fields were left constant.
Fig.~\ref{fig:FvsB} shows the measured value of the beat frequency
between signal and control field as a function of the bias field.
The data points can be well fitted with a linear function, as
shown by the dashed line. Within our experimental accuracy, the
beat frequency between the released signal light and the control
beam frequency corresponds to the frequency of the $\Delta m_F=2$
transition between states $\left|g_-\right\rangle$ and
$\left|g_+\right\rangle$. That is to say, provided that one
remains within the dark resonance window, storage can be done with
a frequency difference somewhat differing from the precise
two-photon resonance with $\delta=0$ (where
$\delta=\omega_s-\omega_c-2g_F\mu_B B$ denotes the two-photon
detuning), the released signal beam frequency is such that the
optical difference frequency then matches the two-photon
resonance. In additional measurements, we have performed the
storage procedure with a variable difference frequency, while the
magnetic bias field was left constant. Fig. \ref{fig:FvsDelta}
shows a measurement of the observed beat frequency between the
released signal beam and the control beam. The shown circles are
the corresponding data points, which have been fitted with a
linear function (solid line). The slope of the fitted line is
$-0.02 (\pm 0.02)$, which within our experimental uncertainty is
consistent with a slope of zero. For detuning values comparable to
the width of the EIT window the amplitude of the released signal
is strongly damped, resulting in a diminished signal-to-noise
ratio. For comparison, the data points shown as squares, which
have been fitted with a linear function shown as a dashed line,
give the measured beat signal of the signal and control beams
before storage.

We now show that the matching effect of the two-photon resonance
upon retrieval can be explained within the dark-state polariton
picture of EIT \cite{PhysRevLett.84.5094}. For simplicity, we
assume that the initial two photon detuning is small so that the
following condition is satisfied \cite{Mewes-PRA-2002}.
\begin{equation}
 \delta \ll \frac{g\sqrt{N}}{\sqrt{\Gamma T}}
\end{equation}
Here $\Gamma$ is the upper state relaxation rate, $g$ is the
probe-field coupling strength, $N$ is the density of atoms and $T$
is the time that the probe pulse propagates inside the medium
before the control field is completely turned off.
Under this condition the absorption of the signal field can be
neglected. Introducing normalized, temporally and spatially slowly
varying amplitudes of the complex signal field, as well as the
control-field Rabi-frequency, $E(z,t) = \sqrt{\hbar
\omega_s/2\epsilon_0} {\cal E}(z,t) {\rm e}^{\{-i\omega_s (t -
z/c)\}}$, and
$\tilde\Omega(z,t) = \Omega\left(t-z \cos\alpha /c\right) {\rm e}^
{\{-i\omega_c (t - z \cos\alpha /c)\}}$,
where $\omega_s$ and $\omega_c$ are the carrier frequencies of
signal and control fields, and  $\alpha$ is the angle between the
propagation axes of signal and control fields, the dark-state
polariton reads
\begin{equation}
 \Psi (z,t) = \cos\theta\, {\cal E}(z,t) - \sin\theta\, \sqrt{N} \rho_{g_+g_-}(z,t).
\end{equation}
Here $\rho_{g_+g_-}(z,t)$ is the temporally and spatially slowly
varying density matrix element in the two ground states
$|g_+\rangle$ and $|g_-\rangle$.  $\theta=\theta(t- z
\cos\alpha/c)$ is the mixing angle determined by the amplitude of
the control field via $\tan\theta = g\sqrt{N}/\Omega(z,t)$, where
$\Omega(z,t) =\Omega(t-z \cos\alpha /c)$.
It is important at this point to keep the retardation of the
control field.
As has been shown in Ref. \cite{Mewes-PRA-2002}, if the two-photon
detuning $\delta=\omega_s-\omega_c-\omega_{g_+g_-}$,
$\omega_{g_+g_-}$ being the two-photon resonance frequency, is
sufficiently small, the propagation equation of the dark-state
polaritons reads
\begin{equation}
\Biggl(\frac{\partial}{\partial t}+c\cos^2\theta(z,t)
\frac{\partial}{\partial z}-i\delta \sin^
2\theta(z,t)\Biggr)\Psi(z,t)=0.
\end{equation}
Solving this equation one finds that an input signal pulse with
two-photon detuning $\delta_0 =\omega_s^0-\omega_c-
\omega_{g_+g_-}$ is mapped onto a stationary spin wave with a
phase oscillating in $z$ direction
with $(\omega_s^0-\omega_c)(1-\cos\alpha)/c$.  Note that if both
fields copropagate, i.e. if $\cos\alpha =1$, the spatial phase of
the stored spin wave actually vanishes. This is because the front
of the control pulse which switches the moving polariton to a
stationary spin wave propagates with the vacuum speed of light and
thus moves in phase with the beat note between pump and probe.
Upon retrieval of the pulse with a control field corresponding to
a mixing angle $\cos^2\theta_{\rm r}$, the polariton oscillates at
a frequency equivalent to a two-photon detuning $\delta$ given by
\begin{equation}
 \delta = \delta_0 (1-\cos\alpha)\cos^2\theta_r\label{eq:delta}
\end{equation}
where it was assumed that the direction of the control field was
not changed. One recognizes that for co-propagating control and
signal pulses, i.e. for $\cos\alpha =1$, the two-photon detuning
of the retrieved pulse vanishes exactly irrespective of its input
value as long as the latter lies within the EIT transparency
window. If control and signal beam propagate in orthogonal
directions the frequency of the retrieved signal pulse is pulled
towards two-photon resonance with the strength of pulling given by
the ratio of the group velocity at read out to the vacuum speed of
light.
The latter can be understood as follows: If the storage generates
a spin wave with nonvanishing $\vec k_{\rm spin}$, i.e. for
$\cos\alpha\ne 1$, phase matching demands that the momentum of the
retrieved probe field fulfills $\vec k = \vec k_{\rm spin} +\vec
k_c $, where $\vec k_c$ is the k-vector of the control field at
read-out. If the group velocity at read out is not equal to $c$,
i.e. if $\cos^2\theta_r\ne 1$, the medium has however an index of
refraction which deviates slightly from unity for a small
two-photon detuning. The latter is responsible for the factor
$\cos^2\theta_r$ in eq.(\ref{eq:delta}).

To conclude, we have shown that light stored in $\Lambda$-type
three-level systems shows a strong pulling of its carrier
frequency towards two-photon resonance upon retrieval. This effect
which holds for an arbitrary two-photon detuning of the input
field within the EIT transparency window is a non-dissipative
effect and can be used to measure the energetic separation of the
ground states without precise knowledge of the input signal
frequency. Within our experimental accuracy, the beat frequency
between the retrieved signal field and the control field does not
depend on the initial two-photon detuning of the pulses before
storage. Since the lower levels of the $\Lambda$ system are in our
case Zeeman sublevels of the Rb ground state the beat frequency
between the released signal and the control beam is expected to
resemble $2g_F\mu_B B$, where $g_F$ denotes the hyperfine g-factor
that equals $1/2$ for our rubidium system. The used method
represents a novel approach for optical (Raman) spectroscopy that
should also be applicable to two-photon transitions with a ladder
level scheme. In preliminary experiments the obtained accuracy of
a magnetic field measurement carried out by recording the spectrum
of a dark resonance was also achieved by evaluating the signal
generated by a single retrieved signal light pulse, but with a
greatly reduced measurement time. We wish to point out that our
light storage spectroscopy should also be applicable to
non-classical light states \cite{appel:093602,Honda2008}. Since
the light storage, despite the frequency conversion, is expected
to preserve the quantum properties of light, when squeezed light
is stored beat frequency measurements of the released light should
be possible with sub-shot-noise precision. This opens up
possibilities for novel quantum limited measurements of atomic
transition frequencies and magnetic fields.

G.N. acknowledges support by the Alexander von Humboldt
Foundation.


\begin{thebibliography}{22}
\expandafter\ifx\csname natexlab\endcsname\relax\def\natexlab#1{#1}\fi
\expandafter\ifx\csname bibnamefont\endcsname\relax
  \def\bibnamefont#1{#1}\fi
\expandafter\ifx\csname bibfnamefont\endcsname\relax
  \def\bibfnamefont#1{#1}\fi
\expandafter\ifx\csname citenamefont\endcsname\relax
  \def\citenamefont#1{#1}\fi
\expandafter\ifx\csname url\endcsname\relax
  \def\url#1{\texttt{#1}}\fi
\expandafter\ifx\csname urlprefix\endcsname\relax\def\urlprefix{URL }\fi
\providecommand{\bibinfo}[2]{#2}
\providecommand{\eprint}[2][]{\url{#2}}

\bibitem[{\citenamefont{Fleischhauer et~al.}(2005)\citenamefont{Fleischhauer,
  Imamoglu, and Marangos}}]{10.1103/RevModPhys.77.633}
\bibinfo{author}{\bibfnamefont{M.}~\bibnamefont{Fleischhauer}},
  \bibinfo{author}{\bibfnamefont{A.}~\bibnamefont{Imamoglu}}, \bibnamefont{and}
  \bibinfo{author}{\bibfnamefont{J.~P.} \bibnamefont{Marangos}},
  \bibinfo{journal}{Rev. Mod. Phys.} \textbf{\bibinfo{volume}{77}},
  \bibinfo{pages}{633} (\bibinfo{year}{2005}).

\bibitem[{\citenamefont{Hau et~al.}(1999)\citenamefont{Hau, Harris, Dutton, and
  Behroozi}}]{Nature.397.594-598(1999)}
\bibinfo{author}{\bibfnamefont{L.~V.} \bibnamefont{Hau}},
  \bibinfo{author}{\bibfnamefont{S.~E.} \bibnamefont{Harris}},
  \bibinfo{author}{\bibfnamefont{Z.}~\bibnamefont{Dutton}}, \bibnamefont{and}
  \bibinfo{author}{\bibfnamefont{C.~H.} \bibnamefont{Behroozi}},
  \bibinfo{journal}{Nature} \textbf{\bibinfo{volume}{397}}, \bibinfo{pages}{594
  } (\bibinfo{year}{1999}).

\bibitem[{\citenamefont{Kash et~al.}(1999)\citenamefont{Kash, Sautenkov,
  Zibrov, Hollberg, Welch, Lukin, Rostovtsev, Fry, and
  Scully}}]{PhysRevLett.82.5229}
\bibinfo{author}{\bibfnamefont{M.~M.} \bibnamefont{Kash}},
  \bibinfo{author}{\bibfnamefont{V.~A.} \bibnamefont{Sautenkov}},
  \bibinfo{author}{\bibfnamefont{A.~S.} \bibnamefont{Zibrov}},
  \bibinfo{author}{\bibfnamefont{L.}~\bibnamefont{Hollberg}},
  \bibinfo{author}{\bibfnamefont{G.~R.} \bibnamefont{Welch}},
  \bibinfo{author}{\bibfnamefont{M.~D.} \bibnamefont{Lukin}},
  \bibinfo{author}{\bibfnamefont{Y.}~\bibnamefont{Rostovtsev}},
  \bibinfo{author}{\bibfnamefont{E.~S.} \bibnamefont{Fry}}, \bibnamefont{and}
  \bibinfo{author}{\bibfnamefont{M.~O.} \bibnamefont{Scully}},
  \bibinfo{journal}{Phys. Rev. Lett.} \textbf{\bibinfo{volume}{82}},
  \bibinfo{pages}{5229} (\bibinfo{year}{1999}).

\bibitem[{\citenamefont{Budker et~al.}(1999)\citenamefont{Budker, Kimball,
  Rochester, and Yashchuk}}]{PhysRevLett.83.1767}
\bibinfo{author}{\bibfnamefont{D.}~\bibnamefont{Budker}},
  \bibinfo{author}{\bibfnamefont{D.~F.} \bibnamefont{Kimball}},
  \bibinfo{author}{\bibfnamefont{S.~M.} \bibnamefont{Rochester}},
  \bibnamefont{and} \bibinfo{author}{\bibfnamefont{V.~V.}
  \bibnamefont{Yashchuk}}, \bibinfo{journal}{Phys. Rev. Lett.}
  \textbf{\bibinfo{volume}{83}}, \bibinfo{pages}{1767} (\bibinfo{year}{1999}).

\bibitem[{\citenamefont{Appel et~al.}(2008)\citenamefont{Appel, Figueroa,
  Korystov, Lobino, and Lvovsky}}]{appel:093602}
\bibinfo{author}{\bibfnamefont{J.}~\bibnamefont{Appel}},
  \bibinfo{author}{\bibfnamefont{E.}~\bibnamefont{Figueroa}},
  \bibinfo{author}{\bibfnamefont{D.}~\bibnamefont{Korystov}},
  \bibinfo{author}{\bibfnamefont{M.}~\bibnamefont{Lobino}}, \bibnamefont{and}
  \bibinfo{author}{\bibfnamefont{A.~I.} \bibnamefont{Lvovsky}},
  \bibinfo{journal}{Phys. Rev. Lett.} \textbf{\bibinfo{volume}{100}},
  \bibinfo{pages}{093602} (\bibinfo{year}{2008}).

\bibitem[{\citenamefont{Liu et~al.}(2001)\citenamefont{Liu, Dutton, Behroozi,
  and Hau}}]{Nature.409.490-493(2001)}
\bibinfo{author}{\bibfnamefont{C.}~\bibnamefont{Liu}},
  \bibinfo{author}{\bibfnamefont{Z.}~\bibnamefont{Dutton}},
  \bibinfo{author}{\bibfnamefont{C.~H.} \bibnamefont{Behroozi}},
  \bibnamefont{and} \bibinfo{author}{\bibfnamefont{L.~V.} \bibnamefont{Hau}},
  \bibinfo{journal}{Nature} \textbf{\bibinfo{volume}{409}}, \bibinfo{pages}{490
  } (\bibinfo{year}{2001}).

\bibitem[{\citenamefont{Phillips et~al.}(2001)\citenamefont{Phillips,
  Fleischhauer, Mair, Walsworth, and Lukin}}]{PhysRevLett.86.783}
\bibinfo{author}{\bibfnamefont{D.~F.} \bibnamefont{Phillips}},
  \bibinfo{author}{\bibfnamefont{A.}~\bibnamefont{Fleischhauer}},
  \bibinfo{author}{\bibfnamefont{A.}~\bibnamefont{Mair}},
  \bibinfo{author}{\bibfnamefont{R.~L.} \bibnamefont{Walsworth}},
  \bibnamefont{and} \bibinfo{author}{\bibfnamefont{M.~D.} \bibnamefont{Lukin}},
  \bibinfo{journal}{Phys. Rev. Lett.} \textbf{\bibinfo{volume}{86}},
  \bibinfo{pages}{783} (\bibinfo{year}{2001}).

\bibitem[{\citenamefont{Eisaman et~al.}(2005)\citenamefont{Eisaman, Andr\'e,
  Massou, Fleischhauer, Zibrov, and Lukin}}]{10.1038/nature04327}
\bibinfo{author}{\bibfnamefont{M.~D.} \bibnamefont{Eisaman}},
  \bibinfo{author}{\bibfnamefont{A.}~\bibnamefont{Andr\'e}},
  \bibinfo{author}{\bibfnamefont{F.}~\bibnamefont{Massou}},
  \bibinfo{author}{\bibfnamefont{M.}~\bibnamefont{Fleischhauer}},
  \bibinfo{author}{\bibfnamefont{A.~S.} \bibnamefont{Zibrov}},
  \bibnamefont{and} \bibinfo{author}{\bibfnamefont{M.~D.} \bibnamefont{Lukin}},
  \bibinfo{journal}{Nature} \textbf{\bibinfo{volume}{438}},
  \bibinfo{pages}{837} (\bibinfo{year}{2005}).

\bibitem[{\citenamefont{Duan et~al.}(2001)\citenamefont{Duan, Cirac, and
  Zoller}}]{Duan2001}
\bibinfo{author}{\bibfnamefont{L.-M.} \bibnamefont{Duan}},
  \bibinfo{author}{\bibfnamefont{J.~I.} \bibnamefont{Cirac}}, \bibnamefont{and}
  \bibinfo{author}{\bibfnamefont{P.}~\bibnamefont{Zoller}},
  \bibinfo{journal}{Science} \textbf{\bibinfo{volume}{292}},
  \bibinfo{pages}{1695} (\bibinfo{year}{2001}), \urlprefix\url{1695}.

\bibitem[{\citenamefont{Chaneliere et~al.}(2005)\citenamefont{Chaneliere,
  Matsukevich, Jenkins, Lan, Kennedy, and Kuzmich}}]{Chaneliere2005}
\bibinfo{author}{\bibfnamefont{T.}~\bibnamefont{Chaneliere}},
  \bibinfo{author}{\bibfnamefont{D.~N.} \bibnamefont{Matsukevich}},
  \bibinfo{author}{\bibfnamefont{S.~D.} \bibnamefont{Jenkins}},
  \bibinfo{author}{\bibfnamefont{S.-Y.} \bibnamefont{Lan}},
  \bibinfo{author}{\bibfnamefont{T.~A.~B.} \bibnamefont{Kennedy}},
  \bibnamefont{and} \bibinfo{author}{\bibfnamefont{A.}~\bibnamefont{Kuzmich}},
  \bibinfo{journal}{Nature} \textbf{\bibinfo{volume}{438}},
  \bibinfo{pages}{833} (\bibinfo{year}{2005}).

\bibitem[{\citenamefont{Vewinger et~al.}(2007)\citenamefont{Vewinger, Appel,
  Figueroa, and Lvovsky}}]{Vewinger2007}
\bibinfo{author}{\bibfnamefont{F.}~\bibnamefont{Vewinger}},
  \bibinfo{author}{\bibfnamefont{J.}~\bibnamefont{Appel}},
  \bibinfo{author}{\bibfnamefont{E.}~\bibnamefont{Figueroa}}, \bibnamefont{and}
  \bibinfo{author}{\bibfnamefont{A.}~\bibnamefont{Lvovsky}},
  \bibinfo{journal}{Opt. Lett.} \textbf{\bibinfo{volume}{32}},
  \bibinfo{pages}{2771} (\bibinfo{year}{2007}).

\bibitem[{\citenamefont{Brandt et~al.}(1997)\citenamefont{Brandt, Nagel,
  Wynands, and Meschede}}]{Brandt1997}
\bibinfo{author}{\bibfnamefont{S.}~\bibnamefont{Brandt}},
  \bibinfo{author}{\bibfnamefont{A.}~\bibnamefont{Nagel}},
  \bibinfo{author}{\bibfnamefont{R.}~\bibnamefont{Wynands}}, \bibnamefont{and}
  \bibinfo{author}{\bibfnamefont{D.}~\bibnamefont{Meschede}},
  \bibinfo{journal}{Phys. Rev. A} \textbf{\bibinfo{volume}{56}},
  \bibinfo{pages}{R1063} (\bibinfo{year}{1997}).

\bibitem[{\citenamefont{Scully and Fleischhauer}(1992)}]{PhysRevLett.69.1360}
\bibinfo{author}{\bibfnamefont{M.~O.} \bibnamefont{Scully}} \bibnamefont{and}
  \bibinfo{author}{\bibfnamefont{M.}~\bibnamefont{Fleischhauer}},
  \bibinfo{journal}{Phys. Rev. Lett.} \textbf{\bibinfo{volume}{69}},
  \bibinfo{pages}{1360} (\bibinfo{year}{1992}).

\bibitem[{\citenamefont{Katsoprinakis et~al.}(2006)\citenamefont{Katsoprinakis,
  Petrosyan, and Kominis}}]{Katsoprinakis2006}
\bibinfo{author}{\bibfnamefont{G.}~\bibnamefont{Katsoprinakis}},
  \bibinfo{author}{\bibfnamefont{D.}~\bibnamefont{Petrosyan}},
  \bibnamefont{and} \bibinfo{author}{\bibfnamefont{I.~K.}
  \bibnamefont{Kominis}}, \bibinfo{journal}{Phys. Rev. Lett.}
  \textbf{\bibinfo{volume}{97}}, \bibinfo{pages}{230801}
  (\bibinfo{year}{2006}).

\bibitem[{\citenamefont{Kominis et~al.}(2003)\citenamefont{Kominis, Kornack,
  Allred, and Romalis}}]{Kominis2003}
\bibinfo{author}{\bibfnamefont{I.~K.} \bibnamefont{Kominis}},
  \bibinfo{author}{\bibfnamefont{T.~W.} \bibnamefont{Kornack}},
  \bibinfo{author}{\bibfnamefont{J.~C.} \bibnamefont{Allred}},
  \bibnamefont{and} \bibinfo{author}{\bibfnamefont{M.~V.}
  \bibnamefont{Romalis}}, \bibinfo{journal}{Nature}
  \textbf{\bibinfo{volume}{422}}, \bibinfo{pages}{596 } (\bibinfo{year}{2003}).

\bibitem[{\citenamefont{Karpa et~al.}(2008)\citenamefont{Karpa, Vewinger, and
  Weitz}}]{Karpa2008a}
\bibinfo{author}{\bibfnamefont{L.}~\bibnamefont{Karpa}},
  \bibinfo{author}{\bibfnamefont{F.}~\bibnamefont{Vewinger}}, \bibnamefont{and}
  \bibinfo{author}{\bibfnamefont{M.}~\bibnamefont{Weitz}},
  \bibinfo{journal}{Phys. Rev. Lett.} \textbf{\bibinfo{volume}{101}},
  \bibinfo{pages}{170406} (\bibinfo{year}{2008}).

\bibitem[{\citenamefont{Fleischhauer and Lukin}(2000)}]{PhysRevLett.84.5094}
\bibinfo{author}{\bibfnamefont{M.}~\bibnamefont{Fleischhauer}}
  \bibnamefont{and} \bibinfo{author}{\bibfnamefont{M.~D.} \bibnamefont{Lukin}},
  \bibinfo{journal}{Phys. Rev. Lett.} \textbf{\bibinfo{volume}{84}},
  \bibinfo{pages}{5094} (\bibinfo{year}{2000}).

\bibitem[{\citenamefont{Mewes and Fleischhauer}(2002)}]{Mewes-PRA-2002}
\bibinfo{author}{\bibfnamefont{C.}~\bibnamefont{Mewes}} \bibnamefont{and}
  \bibinfo{author}{\bibfnamefont{M.}~\bibnamefont{Fleischhauer}},
  \bibinfo{journal}{Phys. Rev. A} \textbf{\bibinfo{volume}{66}},
  \bibinfo{pages}{033820} (\bibinfo{year}{2002}).

\bibitem[{\citenamefont{Lukin et~al.}(2000)\citenamefont{Lukin, Yelin, and
  Fleischhauer}}]{Lukin2000}
\bibinfo{author}{\bibfnamefont{M.~D.} \bibnamefont{Lukin}},
  \bibinfo{author}{\bibfnamefont{S.~F.} \bibnamefont{Yelin}}, \bibnamefont{and}
  \bibinfo{author}{\bibfnamefont{M.}~\bibnamefont{Fleischhauer}},
  \bibinfo{journal}{Phys. Rev. Lett.} \textbf{\bibinfo{volume}{84}},
  \bibinfo{pages}{4232} (\bibinfo{year}{2000}).

\bibitem[{\citenamefont{Karpa and Weitz}(2006)}]{2006NatPh.2.332K}
\bibinfo{author}{\bibfnamefont{L.}~\bibnamefont{Karpa}} \bibnamefont{and}
  \bibinfo{author}{\bibfnamefont{M.}~\bibnamefont{Weitz}},
  \bibinfo{journal}{Nature Physics} \textbf{\bibinfo{volume}{2}},
  \bibinfo{pages}{332} (\bibinfo{year}{2006}).

\bibitem{envelope}
    The shape of the transmitted (initially rectangular shaped) input pulse
    is due to absorption in the dense medium. Upon entry into the cell the atoms are pumped into the dark state before the pulse can propagate lossless.
    The fast ocsillations on the input signal are due to beating of the incident optical fields. The shape of the retrieved signal
  light is determined by the Lorentzian-like spectrum of the transparency
  window, whose Fourier transform is an exponentially decaying function.

\bibitem[{\citenamefont{Honda et~al.}(2008)\citenamefont{Honda, Akamatsu,
  Arikawa, Yokoi, Akiba, Nagatsuka, Tanimura, Furusawa, and
  Kozuma}}]{Honda2008}
\bibinfo{author}{\bibfnamefont{K.}~\bibnamefont{Honda}},
  \bibinfo{author}{\bibfnamefont{D.}~\bibnamefont{Akamatsu}},
  \bibinfo{author}{\bibfnamefont{M.}~\bibnamefont{Arikawa}},
  \bibinfo{author}{\bibfnamefont{Y.}~\bibnamefont{Yokoi}},
  \bibinfo{author}{\bibfnamefont{K.}~\bibnamefont{Akiba}},
  \bibinfo{author}{\bibfnamefont{S.}~\bibnamefont{Nagatsuka}},
  \bibinfo{author}{\bibfnamefont{T.}~\bibnamefont{Tanimura}},
  \bibinfo{author}{\bibfnamefont{A.}~\bibnamefont{Furusawa}}, \bibnamefont{and}
  \bibinfo{author}{\bibfnamefont{M.}~\bibnamefont{Kozuma}},
  \bibinfo{journal}{Phys. Rev. Lett.} \textbf{\bibinfo{volume}{100}},
  \bibinfo{pages}{093601} (\bibinfo{year}{2008}).

\end{thebibliography}
\end{document}